\documentclass[11pt]{amsart}
\usepackage{geometry}                
\usepackage{graphicx}
\usepackage{amssymb}
\usepackage{epstopdf}
\usepackage{natbib}
\usepackage{lineno}
\linenumbers
\usepackage{setspace} 
\title[Wright-Fisher path integral]{A path integral formulation of the Wright-Fisher process with genic selection}
\author{Joshua G. Schraiber}
\date{Started on July 1, 2013; compiled on \today}                                           

\begin{document}
\maketitle

\begin{abstract}
The Wright-Fisher process with selection is an important tool in population genetics theory. Traditional analysis of this process relies on the diffusion approximation. The diffusion approximation is usually studied in a partial differential equations framework. In this paper, I introduce a path integral formalism to study the Wright-Fisher process with selection and use that formalism to obtain a simple perturbation series to approximate the transition density. The perturbation series can be understood in terms of Feynman diagrams, which have a simple probabilistic interpretation in terms of selective events. The perturbation series proves to be an accurate approximation of the transition density for weak selection and is shown to be arbitrarily accurate for any selection coefficient.
\end{abstract}

\section{Introduction}
Modern population genetics theory can be broken down into two broad subclasses: forward-in-time, in which the generation-to-generation allele frequency dynamics are tracked, and backward-in-time, in which genealogical relationships are modeled. While forward-in-time models were developed first, the introduction of the coalescent by \citet{kingman1982coalescent} ushered in a revolution in our understanding of neutral genetic variation. The success of the coalescent in providing a simple framework for analyzing neutral loci has inspired a number of attempts to construct a genealogical representation of models with natural selection \citep{krone1997ancestral, neuhauser1997genealogy, donnelly1999genealogical}. However these models have not been particularly amenable to analysis due to their complicated structure.

The forward-in-time approach remains the most straight-forward method for analyzing genetic variation under the combined effects of genetic drift and natural selection. This approach is characterized by the diffusion approximation to the Wright-Fisher model \citep{ewens2004mathematical}. For many important quantities (such as ultimate fixation probabilities), the diffusion approximation provides a concise, exact analytic expression. These formulas, in terms of common parameters such as the population scaled selection coefficient $\alpha$, allow for an understanding of how different evolutionary forces impact the dynamics of allele frequency change. Assuming a constant population size, exact analytic results from the diffusion approximation can even be used to estimate the distribution of selection coefficients in the genome \citep{boyko2008assessing, torgerson2009evolutionary}.

Unfortunately, when both selection and genetic drift affect allele frequency dynamics, there is no simple analytic expression for the transition density of the diffusion (that is, the probability that an allele currently at frequency $x$ is at frequency $y$ after $t$ time units have passed). Recently, interest in the transition density has been fueled by advances in experimental evolution \citep{kawecki2012experimental} and ancient DNA \citep{wall2012paleopopulation}, leading to the development of numerous methods for estimating the population scaled selection coefficient from allele frequency time series data \citep{bollback2008estimation,malaspinas2012estimating,mathieson2013estimating,feder2013identifying}. Moreover, because the transition density fully characterizes the allele frequency dynamics, many interesting quantities, such as the time-dependent fixation probability, could be calculated once the transition density is known.

While the diffusion approximation allows one to write down a partial differential equation (PDE) that the transition density must satisfy, it has proved challenging to solve in a robust manner either analytically or numerically. Numerical solution of the PDE is, in principle, straightforward by discretization techniques (see \citet{zhao2013complete} for a recent approach that accounts for fixations and losses of alleles). However, because the relative importance of drift and selection depend on the allele frequency, the discretization scheme must be chosen wisely. Another drawback of numerical methods is that they can be quite time consuming; in particular, this is what limits the method of \citet{gutenkunst2009inferring} to 3 populations while using a diffusion approximation to find the site frequency spectrum for demographic inference.

\citet{kimura1955stochastic} provided an analytical solution to the transitional density with selection, in the form of an eigenfunction decomposition with oblate spheroid wave functions. However, he was unable to compute the eigenvalues exactly, instead resorting to perturbation theory. Motivated by the fact that the eigenfunction decomposition of the model with no selection is known, \citet{song2012simple} developed a novel computational method for approximating the transition density analytically. Their method, based on the theory of Hilbert spaces spanned by orthogonal polynomials, is a significant advance and represents the state-of-the-art in terms of finding the transition density with selection. This method still has several limitations, as it needs to be recomputed if a new selection coefficient is chosen; moreover, computation times can be extremely long because they were required to use high-precision arithmetic. 

In this paper, I present a novel method for approximating the transition density of the Wright-Fisher diffusion with genic selection. This method is based on the theory of path integration, which was introduced by \citet{wiener1921average} for Brownian motion and has found substantial success in applications in quantum mechanics \citep{feynman1948space,feynman2012quantum} and quantum field theory \citep{zee2010quantum}. The key insight of of this approach is to associate every path from $x$ at time $0$ to $y$ at time $t$ with a probability, and then integrate over all possible paths to find the transition density. While computing this integral exactly is only possible in the neutral case, I develop a perturbation scheme to approximate it as a power series in $\alpha$ for the case with genic selection. To facilitate computation of this perturbation expansion, I demonstrate the use of a mnemonic, called Feynman diagrams, to compute the transition density to arbitrary accuracy.

\section{Methods}
\subsection{Partial differential equation formulation}
Here I review some preliminaries about the Wright-Fisher diffusion that will prove useful in the following. Denoting by $\phi_\alpha(x,y;t)$ the transition density with genic selection and population-scaled selection coefficient $\alpha$, standard theory shows that $\phi$ satisfies the PDE
\begin{equation}
\frac{\partial}{\partial t}\phi_\alpha(x,y;t) = \frac{1}{2}\frac{\partial^2}{\partial y^2}y(1-y)\phi_\alpha(x,y;t) - \alpha\frac{\partial}{\partial y}y(1-y)\phi_\alpha(x,y;t),
\end{equation}
with the initial condition $\phi_\alpha(x,y,0) = \delta(x-y)$ with $\delta(\cdot)$ the usual Dirac delta function \citep{ewens2004mathematical}.

\citet{kimura1955solution} found that the for the case $\alpha = 0$, the transition density admits an eigenfunction decomposition,
\begin{equation}
4x(1-x)\sum_{i=1}^\infty \frac{2i+1}{i(i+1)}C_{i-1}^{(3/2)}(1-2x)C_{i-1}^{(3/2)}(1-2y)e^{-\frac{1}{2}i(i+1)t},
\end{equation} 
where the $C_i^\lambda(z)$ are the Gegenbauer polynomials.

\subsection{Path integral formulation}
The path integral formulation begins by defining a probability density functional, which assigns a probability density to any path from $x$ to $y$. Then, the total transition probability from $x$ to $y$ is computed by integrating this density over all paths from $x$ to $y$ (Figure 1).

This probability density functional can be developed intuitively by considering the ``short-time transition densities''. Standard theory for diffusion processes shows that for $\delta t \ll 1$, we can approximate
\[
\phi_\alpha(x,y;\delta t) \approx \frac{1}{\sqrt{2\pi x(1-x) \delta t}}\exp\left\{\frac{(y-(x+\alpha x(1-x)))^2}{2x(1-x)\delta t} \right\}
\]

A naive approach might be to attempt to approximate the probability density of a path by diving the interval $[0,t]$ into $n$ intervals of length $\delta t$. Then we approximate with the probability of the so-called ``zig-zag path'',
\[
\mathcal{P}[z] \approx \prod_{i=1}^{n} \phi_\alpha(z_{i-1},z_i;\delta t) dz_i,
\]
with $z_i = z(i \delta t)$. However, this fails for a variety of reasons, in particular the dependence of the diffusion coefficient on the current allele frequency \citep{graham1977path, durr1978onsager}. Instead, I compute the relative probability density function for a path with selection compared to a neutral path. This functional, which can be rigorously derived using Girsanov's theorem \citep{rogers2000diffusions} can be intuitively developed as
\begin{align*}
\mathcal{G}[z] &\approx  \frac{\prod_{i=1}^{n} \phi_\alpha(z_{i-1},z_i;\delta t) dz_i }{\prod_{i=1}^{n} \phi_0(z_{i-1},z_i;\delta t) dz_i } \nonumber \\
&\approx \frac{\prod_{i=1}^n\frac{1}{\sqrt{2\pi z_{i-1}(1-z_{i-1}) \delta t}}\exp\left\{\frac{(z_i-(z_{i-1}+\alpha z_{i-1}(1-z_{i-1})))^2}{2z_{i-1}(1-z_{i-1})\delta t} \right\}dz_i}{\prod_{i=1}^n\frac{1}{\sqrt{2\pi z_{i-1}(1-z_{i-1}) \delta t}}\exp\left\{\frac{(z_i-z_{i-1})^2}{2z_{i-1}(1-z_{i-1})\delta t} \right\}dz_i} \nonumber \\
&= \exp\left\{\alpha \sum_{i=1}^n (z_i-z_{i-1}) - \frac{\alpha^2}{2}\sum_{i=1}^n z_i(1-z_i)\delta t \right\}.
\end{align*}
Thus, as $\delta t \downarrow 0$ and $n \uparrow \infty$ such that $n\delta t = t$, we have 
\begin{equation}
\mathcal{G}[z] = \exp\left\{\alpha(y-x) - \frac{\alpha^2}{2}\int_0^t z(1-z)ds\right\},
\end{equation}
in which the time dependence of $z$ is suppressed for notational convenience. Now, we can write the transition density as the integral over all \emph{neutral} Wright-Fisher paths of the relative probability of that path with selection,
\begin{equation}
\phi_\alpha(x,y;t) = \int_{(0,x)}^{(t,y)}  e^{\alpha(y-x) - \frac{\alpha^2}{2}\int_0^t z(1-z)ds} \mathcal{P}[z]\mathcal{D}z
\label{funcint}
\end{equation}
where $\mathcal{P}[\cdot]$ is the probability functional of a neutral Wright-Fisher path and $\mathcal{D}z$ is a differential on path space. It is important to note that this expression is merely formal, because $\mathcal{P}[\cdot]$ and the differential $\mathcal{D}x$ don't actually exist. A more rigorous approach would require an appeal to Wiener's theory of integrals in path space.

\subsection{Perturbation approximation}
I now show how to approximate the transition density using a perturbation expansion. Note that the first term in the exponential of \eqref{funcint} is independent of the path, and hence we focus on the path integral
\[
\int_{(0,x)}^{(t,y)}  e^{- \frac{\alpha^2}{2}\int_0^t z(1-z)ds} \mathcal{P}[z]\mathcal{D}z.
\]
We begin by expanding the exponential in a Taylor series about $\alpha = 0$,
\begin{align}
\int_{(0,x)}^{(t,y)}  e^{- \frac{\alpha^2}{2}\int_0^t z(1-z)ds} \mathcal{P}[z]\mathcal{D}z &= \int_{(0,x)}^{(t,y)} \sum_{k=0}^\infty (-1)^k\frac{\alpha^{2k}}{2^k}\frac{1}{k!}\left(\int_0^t z(1-z)ds\right)^k \mathcal{P}[z]\mathcal{D}z \nonumber \\
&= \sum_{k=0}^\infty (-1)^k\frac{\alpha^{2k}}{2^k}\frac{1}{k!} \int_{(0,x)}^{(t,y)} \left(\int_0^t z(1-z)ds\right)^k \mathcal{P}[z]\mathcal{D}z.
\label{perturb}
\end{align}
In the Appendix, I show that the exchange of the summation and the integral is justified by Fubini's theorem. This is in stark contrast to the case in quantum physics, in which the exchange of the sum and integral is often not justified, leading to zero radius of convergence in the perturbation parameter.

Thus, the task of approximating the transition density with selection is reduced to the task of computing the functional integrals
\[
\int_{(0,x)}^{(t,y)} \left(\int_0^t z(1-z) \right)^k \mathcal{P}[z]\mathcal{D}z.
\]
Integrals of this form were considered by \citet{nagylaki1974moments} and \citet{watterson1979estimating} although he was focused on the case where the allele is eventually fixed or lost, whereas here we need to consider only those paths that go from $x$ to $y$ in time $t$. To compute these integrals, it is useful to introduce a diagrammatic method, known as a Feynman diagram \citep{feynman2012quantum, chorin2006stochastic}. Borrowing from the language of physics momentarily, we can regard $V(x) = x(1-x)$ as a \emph{potential energy}, and we can consider the allele frequency being \emph{scattered} by the potential.

The idea can be seen in Figure 2. When the integrand is raised to the $k$th power, we imagine that the allele frequency changes neutrally until some time $s_1$, at which point it interacts with the potential and is scattered. Then, it evolves neutral until time $s_2$, at which point it again interacts with the potential and is scattered. This proceeds until the scattering at time $s_k$, after which the allele frequency evolves neutrally to $y$ at time $t$. Because the interaction times $s_i$ could have happened at any time between $0$ and $t$ and the allele frequency $z_i$ at time $s_i$ is random, we integrate over all times and allele frequencies. For example, we can compute
\[
\int_{(0,x)}^{(t,y)} \left(\int_0^t z(1-z) \right) \mathcal{P}[z]\mathcal{D}z = \int_0^t\int_0^1 \phi_0(x,z_1;s_1)z_1(1-z_1)\phi_0(z_1,y;t-s_1)dz_1ds_1
\]
and
\begin{align*}
&\int_{(0,x)}^{(t,y)} \left(\int_0^t z(1-z) \right)^2 \mathcal{P}[z]\mathcal{D}z = \\
&2\int_0^t\int_0^{s_2}\int_0^1\int_0^1 \phi_0(x,z_1;s_1)z_1(1-z_1)\phi_0(z_1,z_2;s_2-s_1)z_2(1-z_2)\phi_0(z_2,y;t-s_2)dz_1dz_2ds_1ds_2,
\end{align*}
where the factor of $2$ comes from the two orderings in which the scatterings happened. In general, the $k$th order Feynman diagram will come with a factor of $k!$ to count the number of orderings of the scattering events.

Because we know the neutral transition density, computing the integrals that arise from Feynman diagrams is straightforward. Unfortunately, the neutral transition density is only known as an infinite series and in practice computing the integrals is more difficult. In the Appendix, I show how to achieve efficient computation of these integrals for arbitrary $k$.

\section{Results}
\subsection{Accuracy of the perturbation expansion}
A simple error bound can be derived for perturbation expansion \citep{harlow2009simple}. For the $k$th-order perturbation expansion, $\phi_\alpha^{(k)}(x,y;t)$, this bound is 
\begin{equation}
\left|\phi_\alpha(x,y;t) - \phi_\alpha^{(k)}(x,y;t) \right| \leq \left|\frac{\alpha^{2(k+1)}}{2^{k+1}(k+1)!}\int_{(0,x)}^{(t,y)} \left(\int_0^t z(1-z)ds.\right)^{k+1} \mathcal{P}[z]\mathcal{D}z \right|.
\end{equation}
As argued in the Appendix, when $t < 4$, this bound is less than
\[
\frac{\alpha^{2(k+1)}t^{k+1}}{8^{k+1} (k+1)!}\phi_0(x,y;t),
\]
which approaches 0 as $k \rightarrow \infty$ for any $\alpha$. Thus, the perturbation expansion convergences to the true transition density for any $\alpha$, provided that $t$ is small enough.

The error bound presented above is rather crude. To get a more informative picture of the accuracy of the perturbation expansion, I compared to simulations. An interesting quantity that sums up the overall accuracy of the perturbation approximation is the time-dependent probability of absorption. This quantity can be calculated analytically as
\[
\int_0^1 \phi_\alpha(x,y;t)dy
\]
and is easily estimated from simulations. The perturbation method proves to be increasing accurate as more terms are added to the expansion (Figure 3). However, even for moderate values of $\alpha$, a large number of terms are required for an accurate approximation. 

\section{Discussion}
The Wright-Fisher process with selection is a primary tool for elucidating the impact of natural selection on genetic variation. However, the transient behavior of the process has been difficult to study, with much work focusing on equilibrium aspects, such as the stationary distribution \citep{wright1931evolution} or the site frequency spectrum \citep{sawyer1992population}. Nonetheless, transient dynamics have an important impact on natural variation and are critical to forming a complete understanding of how natural selection shapes genomes. In this paper, I presented a novel path integral formulation of the Wright-Fisher process with genic selection. This led naturally to a simple perturbation scheme for computing the transition density with weak selection. 

The perturbation expansion of the transition density can be understood by using Feynman diagrams (Figure 1). Although the traditional motivation for Feynman diagrams comes from quantum physics \citep{feynman2012quantum}, they can be interpreted in a population genetic context. For instance, in the first-order term of the perturbation expansion, an allele begins drifting neutrally. At a time when the allele frequency is $z$, there is a probability $\frac{\alpha^2}{2} z(1-z)$ of a selective event occurring, which has a natural interpretation as an individual of one allelic type encountering an individual of the other allelic type, weighted by the strength of selection. After the selective event occurs, the allele again drifts neutrally. Higher-order terms in the perturbation expansion include more selective events. The perturbation expansion is always multiplied by a term that depends only on the difference between initial and final allele frequencies and the selection coefficient. This factor can be thought of as the probability that a one or the other allelic type ``wins'' selective event.

The perturbation scheme described in this paper works best for weak selection. For stronger selection, other approximation methods, such as the Gaussian diffusion approximation \citep{nagylaki1990models,feder2013identifying} or even a deterministic approximation may be more practical. Moreover, the model considered in this paper does not have fully general diploid selection. The path integral approach applies in an equally straightforward fashion to diploid selection, but the mathematics become significantly more complicated. In that case, the orthogonal polynomial method of \citet{song2012simple} may be better suited. 

Path integral formulations have been used successfully in population genetics in the past. \citet{rouhani1987speciation} made use of a path integral to approximate the probability of shifting between selective optima in the context of quantitative trait evolution. Their approximation scheme, however, was quite different than one that I explored. They assumed relatively strong selection and expanded the path integral around the most likely path between the two selective optima, in contrast to the weak selection perturbative approach taken in this paper. More recently, path integrals have been used to examine fitness flux \citep{mustonen2010fitness} and Muller's ratchet \citep{neher2012fluctuations}.

A significant strength of the path integral approach is its adaptability to evolution of a locus with a large number of alleles, each of which corresponds to a phenotypic value. In previous contexts, such a model has been used to model quantitative trait evolution, called a continuum-of-alleles model \citep{kimura1965stochastic}. Earlier approaches to incorporate genetic drift into a continuum-of-alleles model using the theory of measure-valued diffusions \citep{fleming1979some, ethier1987infinitely} have made significant advances in understanding the neutral dynamics of such processes \citep{dawson1982wandering, ethier1993transition, ethier1993fleming, donnelly1996countable}. However, incorporating selection greatly increases the difficulty of obtaining analytical results (but see \citet{donnelly1999genealogical, dawson2001large}). It is possible that a path integral formulation of such a process could lead to a perturbative approach to incorporating selection, in much the same way as the path integral approach has been successful in perturbative quantum field theory \citep{zee2010quantum}.

\section{Software}
A Mathematica program to compute the transition density can be obtained by contacting the author.

\bibliographystyle{plainnat}
\bibliography{references}

\section{Appendix}
\subsection{Exchanging the order of integration and summation in \eqref{perturb} is justified}
To establish this fact, we first need a simple bound on the functional
\[
\mathcal{F}_k[X_s] = \frac{\alpha^{2k}}{2^kk!}\left(\int_0^t z(1-z)ds.\right)^k
\]
Note that, because $z$ represents a frequency, we know that $z$ is bounded between $0$ and $1$ for all $s$. Thus,
\begin{align*}
\int_0^t z(1-z)ds &\leq \int_0^t \frac{1}{2}\left(1-\frac{1}{2}\right)ds \\
&= \frac{1}{4}t.
\end{align*}
Without loss of generality, we can assume that $t < 4$ because if not, we could rescale time (and hence rescale $\alpha$) to ensure that $t < 4$. Therefore,
\[
\mathcal{F}_k[z] \leq \frac{\alpha^{2k}t^k}{8^kk!}.
\]
Now, to apply Fubini's theorem, I must show that
\[
\int_{(0,x)}^{(t,y)} \left( \sum_{k=0}^\infty \mathcal{F}_k[z] \right) \mathcal{P}[z]\mathcal{D}z < \infty
\]
and
\[
\sum_{k=0}^\infty \int_{(0,x)}^{(t,y)} \mathcal{F}_k[z] \mathcal{P}[z]\mathcal{D}z < \infty
\]
For the first case, observe that
\begin{align*}
\int_{(0,x)}^{(t,y)}\left( \sum_{k=0}^\infty \mathcal{F}_k[z]\right) \mathcal{P}[z]\mathcal{D}z &\leq \int_{(0,x)}^{(t,y)} \left(\sum_{k=0}^\infty \frac{\alpha^{2k}t^k}{8^kk!} \right) \mathcal{P}[z]\mathcal{D}z\\
&= \int_{(0,x)}^{(t,y)} e^{\frac{\alpha^2}{8}t} \mathcal{P}[z]\mathcal{D}z \\
&= e^{\frac{\alpha^2t}{8}}\phi_0(x,y;t) \\
&< \infty.
\end{align*}
An extremely similar calculation shows that the second case is true as well.

\subsection{Computation of Feynman diagrams}
The integrals arising from the Feynman diagrams can only be expressed properly as infinite sums. In practice, it is necessary to truncate these sums after a finite number of terms. In this section, I develop an approach efficiently compute the sums. From Kimura's spectral representation of the transition density, it is clear that the $k$th order Feynman diagram results in 
\begin{align}
4x(1-x)\sum_{i_1,i_2,\ldots,i_{k+1}} &C_{i_1-1}^{(3/2)}(1-2x)C_{i_{k+1}-1}^{(3/2)}(1-2y)\prod_{j = 1}^{k+1} \frac{2i_j +1}{i_j(i_j+1)} \nonumber \\
&\times \prod_{j=1}^{k-1} \int_0^1 z_j^2(1-z_j)^2C_{i_j}^{(3/2)}(1-2z_j)C_{i_{j+1}}^{(3/2)}(1-2z_j)dz_j \label{fullexpansion} \\
&\times \int_0^t\int_0^{s_k}\cdots\int_0^{s_1}e^{-(\lambda_{i_1}s_1 + \sum_{j=2}^k \lambda_{i_j}(s_j-s_{j-1}) + \lambda_{i_{k+1}}(t-s_k)}ds_1\cdots ds_{k-1}ds_k, \nonumber
\end{align}
with $\lambda_i = i(i+1)/2$, the eigenvalues of Kimura's transition density.

The integrals over allele frequencies can be done exactly by using the properties of the Gegenbauer polynomials. First,
\begin{equation}
\int_0^1 x(1-x)C_i^{(3/2)}(1-2x)C_j^{(3/2)}(1-2x)dx = -\frac{1}{32}\int_{-1}^1(1-z^2)^2C_i^{(3/2)}(z)C_j^{(3/2)}(z)dz
\label{toughintegral}
\end{equation}
after making the substitution $z = 1-2x$. This puts the Gegenbauer polynomials on their natural domain, $[-1,1]$. Now, multiplying through by one of the factors of $(1-z^2)$,
\[
\int_{-1}^1(1-z^2)^2C_i^{(3/2)}(z)C_j^{(3/2)}(z)dz = \int_{-1}^1(1-z^2)C_i^{(3/2)}(z)C_j^{(3/2)}(z)dz - \int_{-1}^1(1-z^2)zC_i^{(3/2)}(z)zC_j^{(3/2)}(z)dz.
\]
The first term can be recognized as the orthogonality condition for the Gegenbauer polynomials and hence,
\[
\int_{-1}^1(1-z^2)C_i^{(3/2)}(z)C_j^{(3/2)}(z)dz = \delta_{i,j}\frac{2(i+1)(i+2)}{3+2i}.
\]
To simplify the second term, use the recurrence relation for the Gegenbauer polynomials to find that,
\[
zC_i^{(3/2)}(z) = \frac{1}{3+2i}\left((i+1)C_{i+1}^{(3/2)}(z)+(i+2)C_{i-1}^{(3/2)}(z) \right).
\]
Substituting and multiplying through yields
\begin{align*}
\int_{-1}^1(1-z^2)zC_i^{(3/2)}(z)zC_j^{(3/2)}(z)dz = \, &\frac{1}{(2i+3)(2j+3)} \\
& \times \left((i+1)(j+1)\int_{-1}^1(1-z^2)C_{i+1}^{(3/2)}(z)C_{j+1}^{(3/2)}(z)dz \right. \\
&  \quad \times (i+1)(j+2)\int_{-1}^1(1-z^2)C_{i+1}^{(3/2)}(z)C_{j-1}^{(3/2)}(z)dz \\
& \quad \times (i+2)(j+1)\int_{-1}^1(1-z^2)C_{j-1}^{(3/2)}(z)C_{j+1}^{(3/2)}(z)dz \\
& \quad \left. \times (i+2)(j+2)\int_{-1}^1(1-z^2)C_{i-1}^{(3/2)}(z)C_{j-1}^{(3/2)}(z)dz\right).
\end{align*}
Again, these integrals can be simplified using the orthogonality of the Gegenbauer polynomials to finally see that the integral in \eqref{toughintegral} equals
\begin{align}
-\frac{1}{32}&\left(\delta_{i,j}\frac{2(i+1)(i+2)}{3+2i} + \frac{1}{(2i+3)(2j+3)}\left(\delta_{i,j}(i+1)^2\frac{2(i+2)(i+3)}{5+2i} \right. \right. \nonumber \\
& \quad  + \delta_{i,j-2}(i+1)(i+4)\frac{2(i+2)(i+3)}{5+2i} + \delta_{i,j+2}(i+2)(i-1)\frac{i(i+1)}{1+2i} \nonumber \\
& \quad +\left. \left. \delta_{i,j}(i+2)^2\frac{i(i+1)}{1+2i} \right)\right).
\end{align}
An important consequence of this fact is that the many of the terms in the sum \eqref{fullexpansion} are equal to zero.

The integrals over the intermediate times can also be evaluated exactly, although I have not been able to find a general formula. In this case, it is straight-forward to precompute the integral for all possible sets of equal indices and then substitute into the sum.

\newpage
\section{Figures}
\begin{figure}[!htp]
\includegraphics{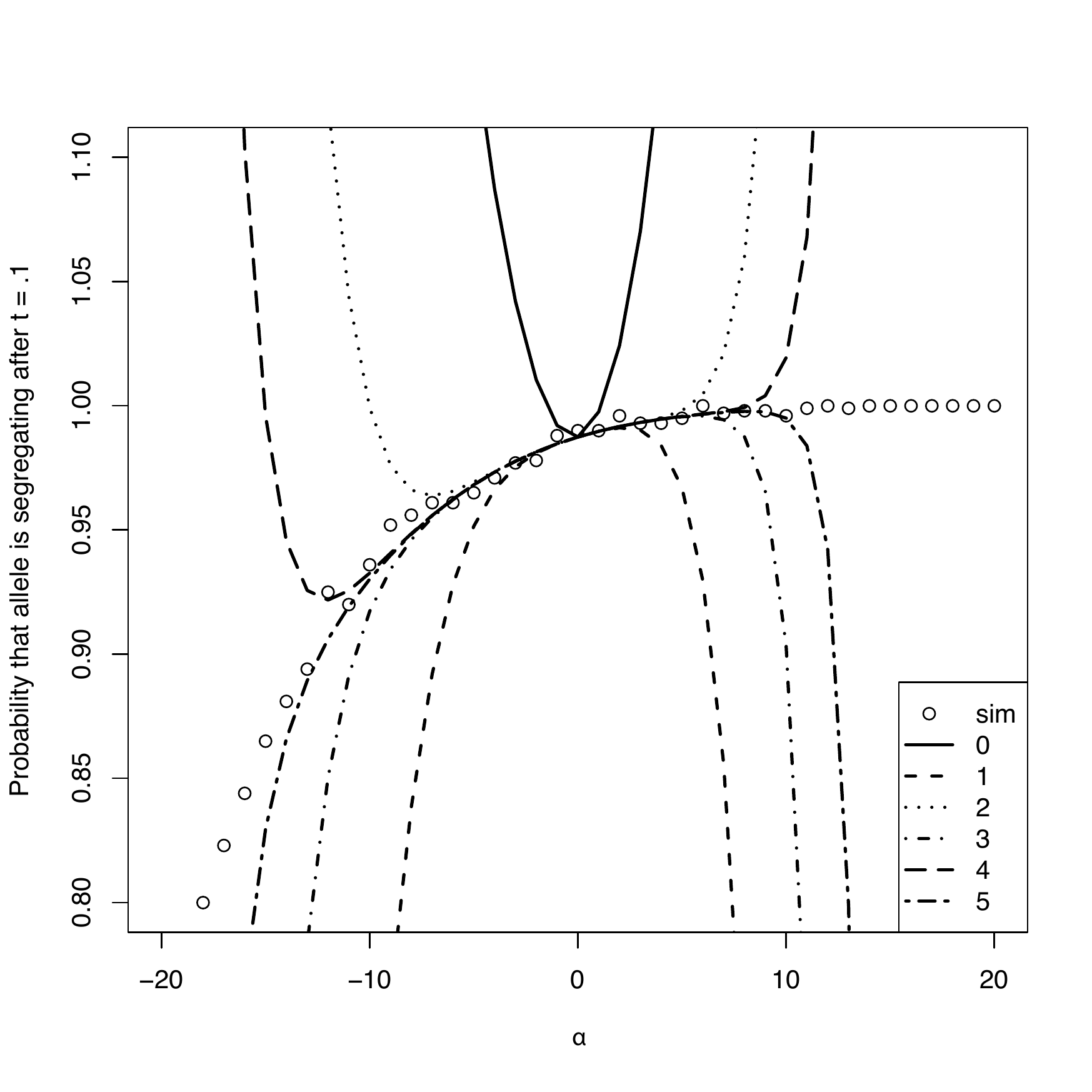}
\caption{Path integrals. Three paths from $x = .2$ to $y = .7$ are depicted. To evaluate a path integral, many such paths are generated and each assigned a probability. Then, the sum of probabilities over all paths is taken. Paths were generated using the method of \citet{schraiber2013analysis}.}
\end{figure}

\newpage
\begin{figure}[!htp]
\includegraphics{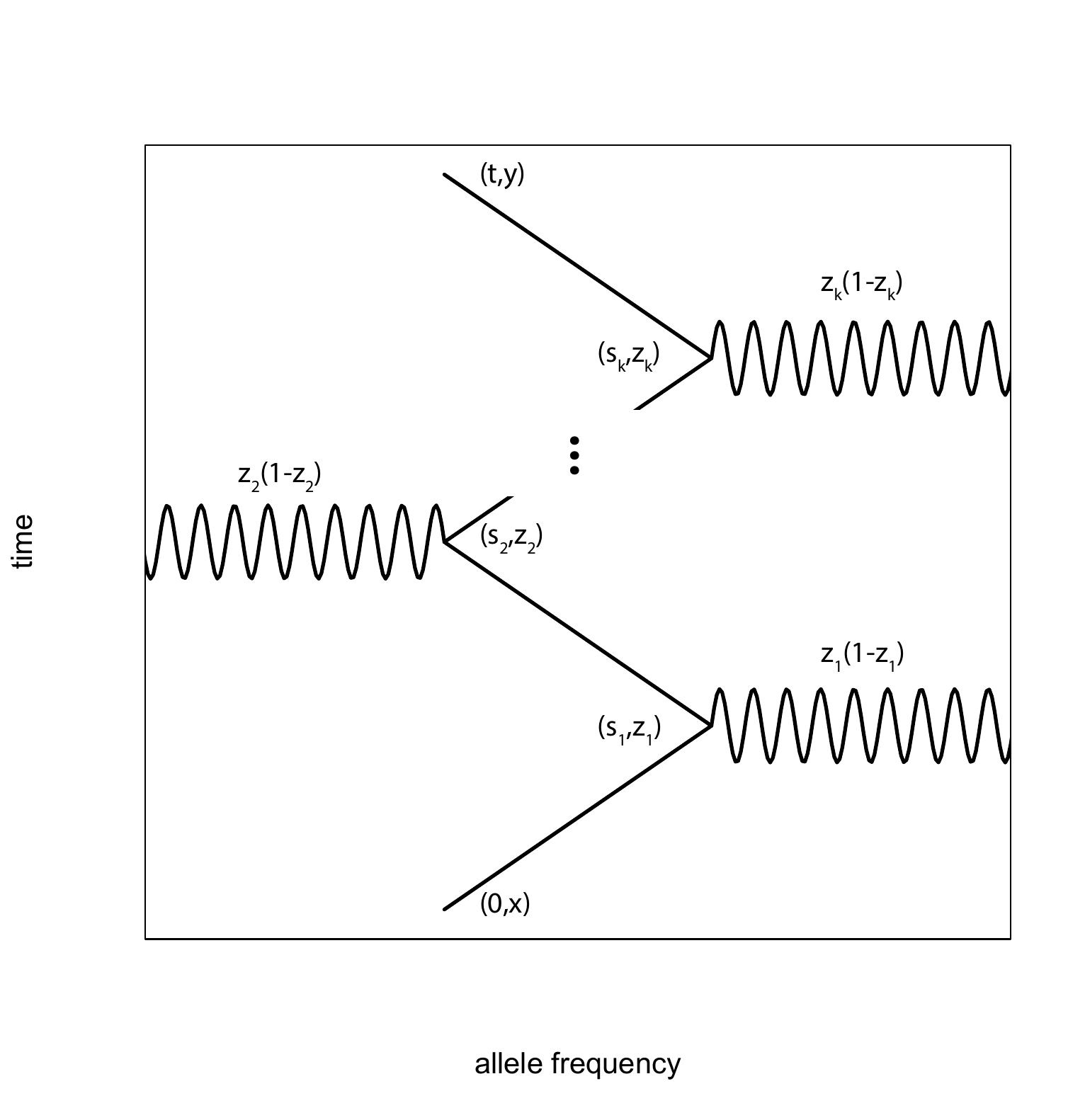}
\caption{Feynman diagrams. Feynman diagrams are used to evaluate the integrals that show up in the perturbation expansion. The allele starts at time $0$ and frequency $x$, evolving neutrally until time $s_1$, when it has frequency $z_1$ and is perturbed by natural selection. It then evolves to time $s_2$ and allele frequency $z_2$, at which point it is again perturbed by natural selection. This continues until the final perturbation at time $s_k$ and frequency $z_k$, after which it evolves neutrally to time $t$ and frequency $y$. }
\end{figure}

\newpage
\begin{figure}[!htp]
\includegraphics{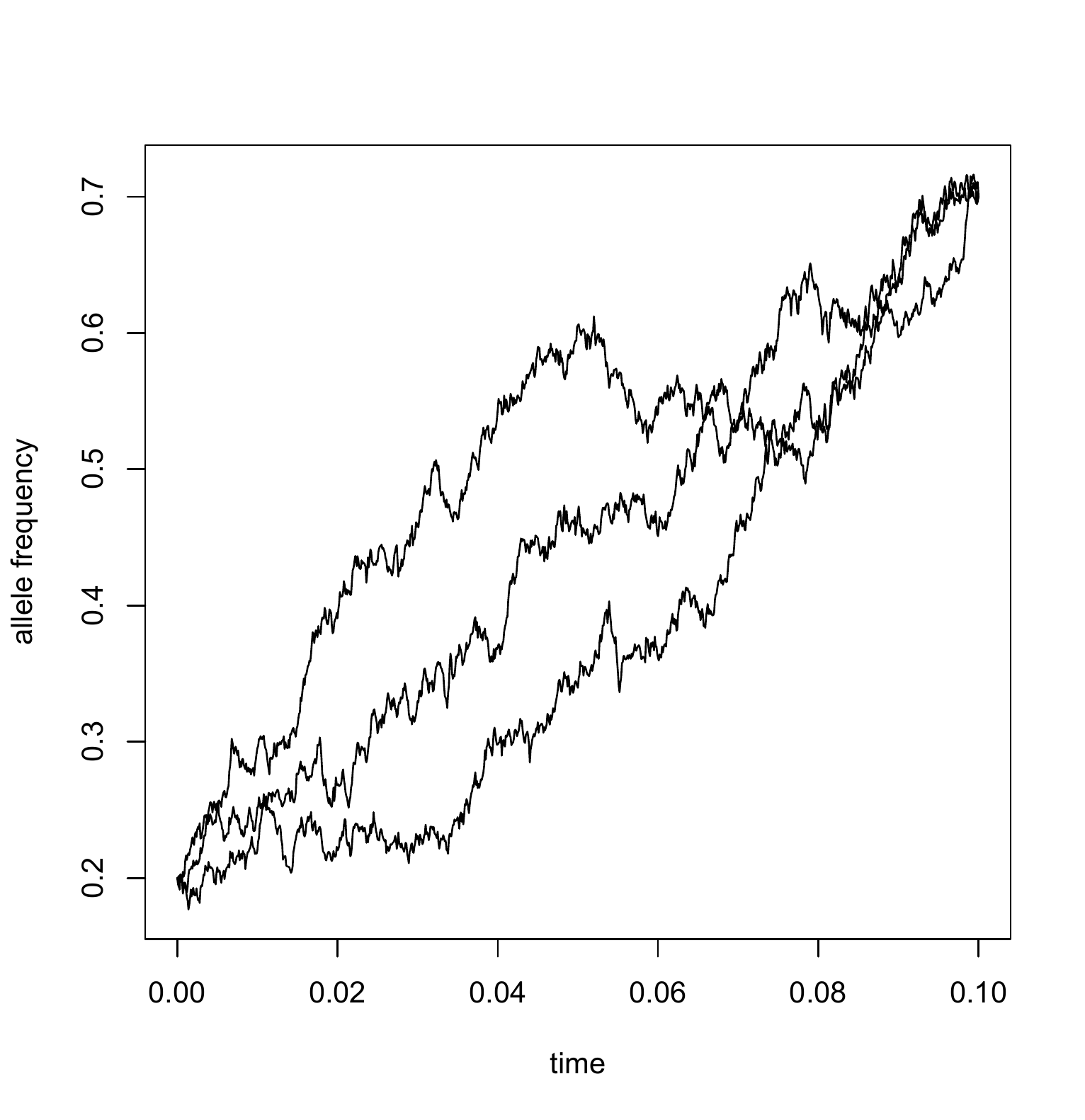}
\caption{Accuracy of the perturbation expansion. The perturbation expansion is compared to simulations of the Wright-Fisher diffusion. An allele with starting frequency $x = .2$ evolves under genetic drift and natural selection for $t = .1$ with a variety of selection coefficients. The probability that the allele was not absorbed is then computed. Dots show the values from simulations while lines indicate successively higher orders of perturbation expansion. }
\end{figure}

\end{document}